\documentclass[12pt]{article}
\usepackage{graphicx} 
\newcommand{\nit}{\noindent}
\newcommand{\no}{\nonumber}
\newcommand{\be}{\begin{equation}}
\newcommand{\ee}{\end{equation}}
\newcommand{\bea}{\begin{eqnarray}}
\newcommand{\eea}{\end{eqnarray}}

\newcommand{\dta}{\mbox{$\delta$}}

\newcommand{\lam}{\mbox{$\lambda$}}
\newcommand{\eps}{\mbox{$\epsilon$}}
\newcommand{\gam}{\mbox{$\gamma$}}

\begin{document}
\thispagestyle{empty}
\title{Time Domain Computation of a Nonlinear Nonlocal Cochlear Model
with Applications to Multitone Interaction in Hearing} 
\author{Jack Xin\thanks{Department of Mathematics and TICAM,
 University of Texas at Austin,
Austin, TX 78712, USA. Corresponding author, email: jxin@math.utexas.edu.
This work was partially supported by ARO grant DAAD 19-00-1-0524.}, 
\hspace{.05 in}Yingyong Qi \thanks{
Qualcomm Inc, 5775 Morehouse Drive, San Diego, CA 92121, USA.},
\hspace{.05 in}  
Li Deng\thanks{Microsoft Research, One Microsoft Way, 
Redmond, WA 98052, USA.}
}
\date{}
\maketitle
\baselineskip=18pt

\begin{abstract}
A nonlinear nonlocal cochlear model of 
the transmission line type is studied in order 
to capture the multitone interactions and 
resulting tonal suppression effects. The model can serve as 
a module for voice signal processing, it is a 
one dimensional (in space) damped dispersive nonlinear PDE based on 
mechanics and phenomenology of hearing. It describes the motion of 
basilar membrane (BM) in the cochlea driven by input 
pressure waves. Both elastic damping and
selective longitudinal fluid damping are present.
The former is nonlinear and nonlocal in BM displacement,
and plays a key role in
capturing tonal interactions. The latter is
active only near the exit boundary (helicotrema), and is
built in to damp out the remaining long waves. 
The initial boundary value problem is numerically solved 
with a semi-implicit second order finite difference method. 
Solutions reach 
a multi-frequency quasi-steady state.
Numerical results are shown on
two tone suppression from both high-frequency and low-frequency 
sides, consistent with 
known behavior of two tone suppression. 
Suppression effects among three tones are 
demonstrated by showing how the response magnitudes of the fixed two tones 
are reduced as we vary the third tone 
in frequency and amplitude. We observe qualitative agreement 
of our model solutions with existing cat auditory neural data.
The model is thus simple and efficient 
as a processing tool for voice signals.  
\end{abstract}
\thispagestyle{empty}
\vspace{.2 in}

\hspace{.1 in} {\bf Communications in Mathematical Sciences, to appear.}
\newpage

\section{Introduction}
\setcounter{equation}{0}
\setcounter{page}{1}
Voice signal processing algorithms have received increasing attention 
in recent years for improving the design of hearing devices and 
for evaluating acoustic theories of peripheral auditory systems, see 
Meddis et al 
\cite{Meddis_01} among others. A fundamental issue a computational method 
needs to resolve is the nonlinear interaction of 
acoustic waves of different frequencies and 
the resulting tonal suppression effects. A nonlinear filter bank 
approach is developed in \cite{Meddis_01}, based on 
knowledge of auditory responses. Nonlinearities are known to originate 
in the cochlea and are further modified in higher level auditory pathways. 
The cochlear mechanics has first principle 
descriptions, and so partial differential equations (PDEs) 
form a natural mathematical framework to initiate computation.   
However, in vivo cochlear dynamics is not a pure mechanical problem, 
and neural couplings are present to modify responses. 
To incorporate both of these aspects, 
we aim to develop a first-principle based PDE approach to voice signal 
processing, where the neural aspect is introduced in the model 
phenomenologically.  

Cochlear modeling has had a long history, and various 
linear models have been studied at length by analytical and 
numerical methods, see Keller and Neu \cite{Keller_85}, 
Leveque, Peskin and Lax \cite{Peskin_88}, Lighthill \cite{Light_81},
 and references therein.  
It has been realized that 
nonlinearity 
is essential for multitone interactions, see
\cite{Hall_77,Kim_86,Boer_96,Geisler_98} 
among others. The nonlinearity could be 
incorporated through micro-mechanics of 
cochlea, 
such as coupling of basilar membrane (BM) to 
inner hair cells \cite{Kanis_Boer_94,Lim_00}. 
Or nonlinearity could be introduced 
phenomenologically based on 
spreading of electrical and neural 
activities between hair cells at different BM 
locations suggested by 
experimental data, see 
Jau and Geisler \cite{Jau_Geisler_83}, Deng \cite{Deng_92}. 
The latter 
treatment turned out to be quite efficient and will be adopted in this 
paper. Their model with 
nonlinear and nonlocal BM damping will be our starting point. 

Multitone interaction requires one to 
perform numerical simulation and analysis 
in the time domain as the resulting 
time dependent problem is strictly speaking 
irreducible to a steady state problem. 
The commonly used cochlear models, 
including those of \cite{Jau_Geisler_83} and \cite{Deng_92}, 
are {\it intrinsically dispersive}. In particular, long waves 
tend to propagate with little decay from entrance point (stapes) 
to the exit  
point (helicotrama). 
Such an issue can be avoided in linear models because a reduction 
to the steady state (or a frequency domain calculation 
using time harmonic solutions) is available
by factoring out 
the time dependence of solutions. The dispersive phenomenon prompted us to 
incorporate a selective 
fluid longitudinal damping term in the model 
of Jau and Geisler \cite{Jau_Geisler_83}, Deng \cite{Deng_92},
operating near the exit point. The role of such a term 
is to suck out the long waves accumulating near the exit.  
Selective positive or negative damping has been  
a novel way to filter images in PDE method of image processing, see 
Osher and Rudin \cite{Osher_90} for the first work in this direction. 
Our treatment here is similar in spirit though on a different 
type of PDE. 

We then present 
a second order semi-implicit finite 
difference method of our model. The method relaxes 
the stability time step constraint of explicit methods without going into 
the complexity of fully implicit schemes. As our model is nonlocal and 
nonlinear, semi-implicit scheme is a reasonable option for 
computation. 
Numerical examples showed that the method is efficient, 
and that the selective damping term indeed 
removes the long waves at the exit boundary point for a time 
domain calculation of a single tone. The examples also 
demonstrated that the analysis of dispersive waves and slow decay 
of long waves is robust, and the phenomenon persists in the 
nonlinear nonlocal regime of the model. 
 
Subsequently, we show computational results on isodisplacement 
curves for five characteristic frequencies consistent with 
those of Neely \cite{Neely_85} for BM displacement of 1 nm. In case of 
two tone interactions, our model gives both low and high side 
suppression, in qualitative agreement with similar studies of Geisler
\cite{Geisler_98}. For understanding three tone interactions, we 
fix the frequencies and amplitudes of two tones, and vary the 
third tone both in frequency and amplitude in a manner similar 
to auditory neural experiment of Deng and Geisler \cite{Deng_Geisler_87}. Again, our model
showed suppressing effect on the two tones by the third tone, in 
qualitative agreement with experimental data. 
To our best knowledge, the current paper is the first on time
domain simulation of
interactions of three tones using first principle based
PDEs.
The stage is 
set for further computation, analysis, and model development
on more complicated tones.

The rest of the paper is organized as follows.
In section 2, we present our model and its background, also analysis 
of dispersive properties of solutions. In section 3, we outline 
our semi-implicit discretization of the model. In section 4, 
we give model parameters, and show numerical results on 
dispersion of model solutions, isodisplacement curves, 
two and three tone interactions in comparison with existing data. 
The conclusions are in section 5.

\section{Nonlinear Nonlocal Cochlear Model}
\setcounter{equation}{0}
\subsection{Background and Model Setup}
The cochlear modeling has had a long history 
driven by advancement of cochlear 
experiments, see \cite{Bekesy_60,Rhode_71,Allen_80,Geisler_98} among others. 
A brief derivation of the 
one dimensional (1-D) cochlear model of transmission line (long wave) type,  
based on fluid mechanics and elasticity, can be found in Sondhi \cite{Sondhi_80}, see also 
references therein for earlier contributions, and  
de Boer \cite{Boer_96} for later development and higher dimensional models. 
A general form of the 1-D model can be expressed as:
\bea
  & & p_{xx} - N u_{tt} = \epsilon_{s} u_{t}, \;\; x \in (0,L), \label{C0} 
  \\ & & p= m u_{tt} + r(x,u) u_{t} + s(x) u, \label{C1}
\eea
where $p$ is the fluid pressure difference across the basilar membrane (BM), 
$u$ the BM displacement, $L$ the longitudinal length of BM; stapes is at $x=0$, 
and helicotrema is at $x=L$; $N$ a constant depending on fluid density and cochlear 
channel size; $\epsilon_s$ depends on the damping of longitudinal fluid motion 
\cite{Lut_86}; $m$, $r$, $s$ are the mass, damping, and stiffness of BM per unit 
area, with $m$ a constant, $s$ a known function of $x$, 
and $r$ a function(al) of $x$, $u$. 

Choosing a functional form for $r$ is a key step in model design. 
Many choices in the literature assume that $r$ is a local nonlinear function, 
\cite{Hall_77,Diep_etal_87,Wada_93,Kanis_Boer_94,Boer_96}. The nonlinearity 
is necessary to model two tone nonlinear interaction (e.g. two tone suppression effects).
However, as pointed out by Kim \cite{Kim_86}, local form of $r$ is not sufficient to 
account for two tone suppression when suppressors are below the excitors in 
frequency (low side), unless a ''second filter'' is incorporated, see 
e.g  \cite{Hall_77}. 
To achieve the second filter effect on more physical grounds, 
micromechanical approaches have been developed to 
derive damping $r$ from motion of outer hair cells and other physiologically 
relevant parts of cochlear, \cite{Kim_86,Wada_93,Boer_96,Lim_00}
 among others. Though this is a reasonable approach, one 
faces the daunting complexity of the cochlear micromechanics,  
 see \cite{Lim_00} 
for recent progress along this line. 
The alternative approach, first 
proposed in this context by Jau and Geisler \cite{Jau_Geisler_83}, is 
phenomenological in nature. It hypothesized that there is 
longitudinal coupling in the cochlear partition itself, and damping $r$ is 
a convolution integral of $u$ with an exponential weight function on BM. 
Similar idea appeared also in Lyon \cite{Lyon_82} in coupled automatic gain control framework. 
Possible physiological mechanisms underlying the spatially coupled nonlocal damping 
were identified in Deng \cite{Deng_92}, and the resulting model was then used to process 
acoustic signals. 

We shall adopt the latter nonlocal damping approach for choosing the function $r$.
The BM damping coefficient is specified as in  Jau and Geisler \cite{Jau_Geisler_83}, 
Deng \cite{Deng_92}:
\be
r(x,t,u)= r_0(x) +\gam \, \int_{0}^{L}\, |u(x',t)|\exp\{ -|x -x'|/\lam\}\, dx', \label{C2}
\ee
where $r_0(x)$ is the passive BM damping; $\gam $ and $\lam $ are positive 
constants.  The size of $r$ controls how sharp the BM traveling wave envelope 
would be. 

We shall take $\eps_s$ as a function of 
$x$ so that $\epsilon_s =\epsilon_s(x) \geq 0$ 
is a smooth function and is supported 
near the vicinity of $x=L$, for instance:
\be
\eps_s (x)= {\eps \over 1 + \exp\{\beta (x_s -x)\}},\,\, \label{C3}
\ee
where $x_s$ is a constant 
slightly smaller than $L$; $\eps > 1$, $\beta > 1$, are 
two constants, with $\beta $ suitably large.
Taking $\eps_s $ as a spatially dependent function 
can be considered as a way to 
{\it selectively introduce damping} so that possible 
low frequency waves near $x=L$ are damped out and there is 
minimal wave accumulation (or reflection) 
close to the helicotrema (at $x=L$). 
This turns out to be an essential stabilizing mechanism for our time domain 
numerical computation. We will come back to this point later. 
  
The coefficients $m$, $n$, and $s(x)$ are standard. The function 
$s(x)$ is based on the data of Liberman \cite{Lib_82}: 
\[ s(x) = 4 \pi^{2} \, m \, (0.456\, \exp(\, 4.83\,  (\, 1-x/3.5\, ))-0.45\, )^2. \]

The physical boundary and initial 
conditions are:
\bea
& & p_{x}(0,t)= T_{M} p_{T}(t) ,\;\;  p(L,t)=0, \label{C4} \\
& & u(x,0)=u_{t}(x,0) = 0, \label{C5}
\eea
where $p_{T}(t)$ is the input sound pressure at the eardrum; and $T_M$ is a 
bounded linear operator on the space of bounded continuous functions, 
with output depending on the frequency content of $p_T(t)$. 
If $p_T = \sum_{j=1}^{J_M} A_j \exp\{i\omega_j t\} + c.c.$, 
a multitone input, c.c denoting complex conjugate, $J_M$ a positive 
integer, 
then $T_M p_T(t) =\sum_{j=1}^{J_M} B_j \exp\{i\omega_j t\}+ c.c$, where 
$B_j = a_{M}(\omega_j) A_j$, c.c for complex conjugate, 
$a_M (\cdot) $ a scaling function built 
from the filtering characteristics of the  
middle ear. Established data are available in Guinan and Peake
\cite{Guin_Peake_66}. The model setup is now complete. 

\subsection{Model Properties and Analysis}
We briefly go over some special regimes and solutions of the model system to illustrate 
the basic mathematical properties and issues. 
If $m=0$, $r=0$, $\eps=0$, (\ref{C0})-(\ref{C1}) reduces 
to a second order wave equation: $nu_{tt} - (s(x)u)_{xx} = 0$. When $\eps$ is turned on, 
and $\beta =0$, we have a damped second order wave equation driven by boundary condition at $x=0$. 
Now suppose that all coefficients are constants,  then (\ref{C0})-(\ref{C1}) 
considered on the entire line admits planar wave solutions of the form:
\[u = u_0 \exp\{i( k x - \omega t)\}, \, p = p_0 \exp\{i( k x - \omega t) \}, \]
where $k$ is the spatial wave number, $\omega$ the temporal frequency;
 $u_0$ and $p_0$ are respectively the amplitude of displacement and 
pressure. Upon substitution, it follows that:
\bea
& & p_0 = (-m \omega^2 -i r \omega + s) u_0, \no \\
& & (k^2 m + N)\omega^2 + i(k^2 r + \eps) \omega - sk^2 = 0. \label{C6}
\eea

If the damping coefficients $r=\eps = 0$, then:
\be
\omega = \pm {k s_{0}^{1/2} \over (k^2 m + N)^{1/2}}, \label{C7}
\ee
showing that the system is {\it dispersive}, see \cite{Whith_79}, 
i.e, $\omega =\omega (k)$ and $\omega''(k) \not \equiv 0$. 
Waves of different wave length ($2\pi/k$) travel at 
different velocities. The dispersion relation (\ref{C7}) 
gives the group velocity:
\be
\omega'(k) = \pm s_{0}^{1/2}N (m k^2 + N)^{-3/2}, \label{C8}
\ee
which decays to zero as $k \to \infty$. In other words, {\it short waves ($k$ large) 
do not disperse as fast as long waves ($k$ small) }. 

When damping coefficients $r$, $\eps > 0$ ($\beta =0$), we have a damped dispersive system, 
(\ref{C7}) is modified into:
\be
\omega = { -i(k^2 r +\eps)\pm \sqrt{4s_0 k^2 (k^2 m+N) -(k^2r+\eps)^2} \over 2(k^2 m+N)},
\label{C9}
\ee
with $Im \{\omega\} = -(k^2 r +\eps_s)/2(k^2 m + N) < 0$, if
the discriminant is nonnegative; otherwise:
\[ Im \{\omega\} = - { (k^2 r +\eps_s)\pm \sqrt{-4s k^2 (k^2 m+N)
+(k^2r+\eps_s)^2} \over 2
(k^2 m+N)}  < 0. \]
In either case, we see
decay of
planar waves. The decay rate $Im\{\omega\}$ shows
that if $\eps_s =0$, {\it the decay of
long waves ($|k| \ll 1$) is very slow. } Hence the decay
of long waves relies on $\eps_s$.

We remark that the cochlear model in \cite{Deng_92,Deng_K_93}
also contains a so called stiffness coupling term in equation (\ref{C1}), so the 
elastic response becomes:
$
p= m u_{tt} + r(x,t,u) u_{t} + s(x)u -K(x) u_{xx}, 
$
for some positive function $K(x)$, rendering the highest spatial derivative 
order four in the model. Nevertheless, it is easy to check that 
the additional term does not change the dispersive nature of model solutions, 
neither the long wave {\it dispersion and slow decay} properties.

As we will see, the long wave {\it dispersion and 
slow decay} properties persist when coefficients 
become variables, and could cause problems for numerics. However, 
the generic dispersive 
effect of the time dependent solutions is absent in the special case 
when the temporal dependence of solution can be factored out. 
This is the case when nonlinearity is absent, i.e. $r=r(x)$. Let the input 
$P_T(t)$ be a single tone, or in complex variables 
$P_T(t) = A \exp\{ i\omega t\} + c.c$. 
Writing solution as $u= U(x) \exp\{i\omega t\} + c.c.$, 
$p= P(x)\exp\{i\omega t\}+ c.c$, we derive 
the following boundary value problem:
\bea
& & P(x)= (-m\omega^2 + i r(x)\omega + s(x)) U, \label{C10} \\
& & P_{xx} + (N\omega^2 -\eps(x) i \omega)U=0, \label{C11}
\eea
subject to the boundary conditions: 
\be
P_x (0)= T_M A, \, P(L)=0. \label{C12}
\ee
This is the way frequency domain calculation is done, turning a time 
dependent problem into a steady state problem on wave amplitudes, avoiding 
transients and the dispersion effect. However, when nonlinearity is 
present, it is in general  
impossible to factor out the time dependence exactly, especially 
when the input signal consists of multiple tones. 
To do an accurate computation directly on system (\ref{C0})-(\ref{C5}), 
one has to proceed in the time domain, or solving the initial boundary value 
problem in time until transients die out and solutions approach a limiting 
time dependent state (quasiperiodic in general for multiple tones).

\section{Numerical Method}
\setcounter{equation}{0}
In this section, we discuss our numerical method for 
solving system (\ref{C0})-(\ref{C5})
and related numerical issues. 
Solutions of cochlear model with tonal inputs behave like traveling waves 
during early time, and eventually settle down to time periodic states of different periods
at various BM locations according to the resonance information 
(frequency to place mapping) encoded in the function 
$s(x)$. A time domain computation will provide the stabilized limiting multi-frequency 
oscillating states after a long enough time integration. An ideal numerical 
method would be one that can reach the limiting states with accuracy, low cost and 
speed. 

Let us first briefly review some existing methods. 
For a cochlear model with locally nonlinear damping function $r$
(van der Pol type), a method of line discretization is implemented 
in \cite{Diep_etal_87}. Discretization in space is a Galerkin finite 
element method (of second order, with piecewise linear basis), and time stepping is 
explicit variable step 4th order Runge-Kutta, found to be superior to earlier explicit
time stepping schemes such as Heun's and modified Sielecki methods. A fully 
discrete explicit finite difference discretization, first order in time
and second order in space, 
is carried out in \cite{Deng_92,Deng_K_93}. The discretization 
can be viewed as central differencing in space and forward Euler in time. The explicit 
nature of these methods put a severe stability constraint on the time step, requiring long 
computation time to reach stable quasi-periodic states that we are most interested in. 
Also it is known that one step explicit methods 
such as Runge-Kutta, Heun and Euler, 
are prone to accumulation of truncation errors for approximating dynamic 
objects like limit cycles, see \cite{Golub_Ort79}, chapter 2. 

Though implicit methods typically relax the stability 
constraints on time step and are better for computing steady states, 
they are also likely to be slow due to Newton iteration at each time step, and 
the non-banded complicated Jacobian matrix as a result of the nonlocal damping term. 
When $r$ is only Lipschitz continuous in $u$, lack of smoothness is another concern 
for the condition of the associated Jacobian matrix.  
This motivates us to look into semi-implicit methods that allow larger time 
steps and better stability properties than explicit methods, 
while being cheaper and faster than fully implicit schemes.  

Let us introduce our 2nd order accurate semi-implicit discretization. 
We denote by $u_{j}^{n}$ ($p_{j}^{n}$) the numerical approximation of solution 
$u(jh, nk)$ ($p(jh,nk)$), where $h$ the grid size for $x \in [0,L]$, $k$ the time step. 
Let $\dta_{\pm,h}$ be the forward/backward finite differencing operator in $x$, 
$\dta_{\pm,k}$ the forward/backward finite differencing operator in $t$. 
Central first differencing operator in space is $\dta_{0,h}=
{\dta_{+,h} +\dta_{-,h} \over 2}$, and central second differencing operator 
in space is $\dta^{2}_{h}= (\dta_{+,h} -\dta_{-,h})/h$; with similar 
notations $\dta_{0,k}$, $\dta^{2}_{k}$ for 
central differencing operator in time. 
The semi-implicit method for system (\ref{C0}-\ref{C5}) is:
\bea
& & N \dta^{2}_{k} u^{n}_{j} = {1\over 4} \dta_{x}^{2} 
(p^{n+1}_{j} + 2 p^{n}_{j} + p^{n-1}_{j}) - \eps_{j} 
\dta_{0,k}u^{n}_{j},  \label{C13}\\
& & p^{n+1}_{j} 
= m{2 u^{n+1}_{j} - 5 u^{n}_{j} + 4u^{n-1}_{j} - u^{n-2}_{j} \over k^2}
 + r^{n}_{j} (u_t)^{n+1}_{j} + s_{j} u^{n+1}_{j}, \label{C13a} 
\eea
where $1\leq j \leq J$, $n \geq 2$, $\eps_j = \eps_{s} (x_j)$, and:
\be
r_{j}^{n} =  [r_{0,j} + \gam h ({1\over 2} 
|u^{n+1}_{1}| e^{-|x_{j} -x_{1}|/\lam} 
+ {1\over 2} |u^{n+1}_{J}| e^{-|x_{j} -x_{J}|/\lam} 
+ \sum_{l=2}^{J-1} \, |u^{n+1}_{l}|e^{-|x_{j} -x_{l}|/\lam})], \label{C14}
\ee
where $(u_t)^{n+1}_{j}$ is approximated at second order by:
\bea
(u_{t})^{n+1}_{j} & = & (u_{t})^{n-1}_{j} + 
(2k)(u_{tt})^{n-1}_{j} + O(k^2) \no \\
& = & \dta_{0,k} u^{n-1}_{j} + {2k \over N}((p_{xx})_{j}^{n} - \eps_{j} (u_{t})_{j}^{n-1}) 
+O(k^2) \no \\
& = & \dta_{0,k} u^{n-1}_{j} + {2k\over N}(\dta_{h}^{2} p^{n-1}_{j} 
- \eps_{j} \dta_{0,k} u^{n-1}_{j}) + O(k^2), \label{C15}
\eea
and:
\bea
|u^{n+1}_{l}| & = & | u^{n}_{l} + k (u_{t})_{l}^{n} + O(k^2) | \no \\
 & = & |u^{n}_{l} + (4u^{n-1}_{l} - u^{n-2}_{l} 
- 3u^{n}_{l})/(-2)| + O(k^2) \no \\
& = & |(4u^{n-1}_{l} - u^{n-2}_{l} - 5u^{n}_{l})/(-2)|+ O(k^2). \label{C16}
\eea
In (\ref{C15}), we have used the equation (\ref{C0}) once to lower time derivative 
of $u$ by one order. In (\ref{C13a}), 
(\ref{C14}) and (\ref{C16}), we have also used 
one sided second order differencing to approximate $u_{tt}$, $u_{t}$, and 
trapezoidal rule to approximate the spatial integral. 
The variables at $t=(n+1)k$ are all linear. If follows from (\ref{C13a}) that:
\be
u^{n+1}_{j} = { k^2 p^{n+1}_{j} - m (-5 u^{n}_{j} +4 u^{n-1}_{j} - u^{n-2}_{j}) 
+ k r^{n}_{j}(4 u^{n}_{j} - u^{n-1}_{j})/2 \over 
2m + 3k r^{n}_{j}/2 + k^2 s_{j} }. \label{C17}
\ee
Substituting (\ref{C17}) into (\ref{C13}) to eliminate $u^{n+1}$, we get 
the following linear system of equations on $p^{n+1}$ ($\lam = k/h$):
\bea
& & [-(N+{\eps_{j} k\over 2})k^2/(2m +{3k r^{n}_{j}\over 2} + k^2 s_j) 
-{\lam^2\over 2}]p^{n+1}_{j} 
+{\lam^2 p^{n+1}_{j+1}\over 4} + {\lam^2 p^{n+1}_{j-1}\over 4} \no \\
& = & N(-2u^{n}_{j} +u^{n-1}_{j}) + (N+{\eps_{j} k\over 2})
{ - m(-5 u^{n}_{j} +4u^{n-1}_{j} - u^{n-2}_{j}) + {k r^{n}_{j}\over 2}
(4 u^{n}_{j} - u^{n-1}_{j}) \over (2m + {3k r^{n}_{j}\over 2} + k^2 s_{j}) } \no \\
& & -{\lam^2 \over 4}[2(p^{n}_{j+1} -2 p^{n}_{j}+p^{n}_{j-1})+p^{n-1}_{j+1} 
-2p^{n-1}_{j} + p^{n-1}_{j-1}] -{\eps_{j} k\over 2} u^{n-1}_{j}, \label{C18}
\eea
which is solved by inverting a diagonally dominant tridiagonal matrix. Once 
$p^{n+1}$ is computed, $u^{n+1}$ is updated from (\ref{C17}). 

The first two time steps are initiated as follows. Initial condition, 
(\ref{C0}), and (\ref{C1}) give:
\[ p_{xx} - {N\over m} p =0,\;\; p_x(0)=f(0),\, p(L,0)=0, \]
whose solution is:
\be
p(x,0)={f(0)\over \sqrt{N/m} (1 +e^{-2L\sqrt{N/m}})}( \exp\{(x -2L)\sqrt{N/m}\} 
-\exp\{ -\sqrt{N/m}\, x\}). \label{C19}
\ee
At $t=k$, we found 
\[ u(x,k)= k^{2} u_{tt}/2 + O(k^3) = {k^2\over 2m}p(x,0) + O(k^3); \]
and similarly:
\[ u_{t}(x,k)= k p(x,0)/m +k^2 u_{ttt}(x,0) /2 + O(k^3). \]
To find $p(x,k)$, denote $q(x)=p_{t}(x,0)= m u_{ttt}(x,0)+r(x,0)u_{tt}(x,0)$. 
It follows from (\ref{C0}) that $q_{xx} - N u_{ttt}(x,0) =\eps (x) u_{tt}(x,0)$, implying 
\[q_{xx} - {N\over m}q = -\eps (x) u_{tt}(x,0)- {N\over m} r(x,0)u_{tt}(x,0), \]
or a uniquely solvable two point boundary value problem on $q$:
\be
q_{xx} -{N\over m} q = -( N r(x,0)/m^2 - \eps (x))p(x,0), \label{C20}
\ee
with boundary condition: $q_{x}(0)=f'(0)$, $q(L)=0$. 

It follows then from the definition of $q$ that:
\bea
 u_{ttt}(x,0) & = & q(x)/m - r(x,0)p(x,0)/m^2, \no \\
u_{t}(x,k) & = & kp(x,0)/m + {k^{2}\over 2}(q(x)/m - r(x,0)p(x,0)/m^2). \label{C21}
\eea
Equations (\ref{C0})-(\ref{C1}) at $t=k$ implies the uniquely solvable two point boundary 
value problem on $p(x,k)$:
\be
p_{xx} - {N \over m} p = (\eps (x) - N r(x,u))u_{t} - N s(x) u/m, \label{C22}
\ee
with boundary condition: $p_{x}(0)=f(k)$, $p(L)=0$. Both (\ref{C20}) and 
(\ref{C22}) are solved with a standard second order discretization. We then 
will have computed $(u,p)(x,k)$. Similarly, with $k$ replaced by $2k$, we 
compute $(u,p)(x,2k)$.

The left boundary condition $p_{x}(x,0)=f(t)$ is discretized as: 
$p^{n+1}_{2} - p^{n+1}_{0} = 2h f((n+1)k)$, with second order accuracy. The right 
boundary condition $p^{n+1}_{J+1} = 0$ is exact. 
The semi-implicit method is now completely defined, and it reduces to an unconditionally 
stable implicit method for standard second order 
wave equation when $m=0$, $s$ is constant, and damping is absent, \cite{Strik_89}, 
chapter 8. 

\section{Numerical Results and Multitone Interaction}
\setcounter{equation}{0}
In this section, we present numerical results based on our model and numerical methods 
discussed in the last two sections. First we list all parameters of our cochlear model
and those of the associated middle ear model based on 
Guinan and Peake \cite{Guin_Peake_66}. We then
show computed single tones and demonstrate the dispersion
effects of long waves and the role of selective longitudinal fluid damping in our model.
As a test of model tuning property, we also show the computed isodisplacement 
curves (at characteristic frequencies 500 Hz, 2 kHz, 4 kHz, 6 kHz, 10 kHz). 
Subsequently, we give numerical examples of both 
the high side and low side suppression of two tone interaction in model solutions
consistent with Figures 10.2, 10.4, 10.7 in \cite{Geisler_98}. The asymmetry of 
high and low side suppression is captured by the model. We then show numerical 
simulation of three tone interaction, presenting results on how the amplitudes of 
the two of the three tones change as we vary the amplitude and frequency of the third 
tone, qualitatively consistent with experimental data of 
Deng and Geisler \cite{Deng_Geisler_87} 
on responses of auditory 
neural fibers to input of three tones.

\subsection{Model Parameters}
The parameters for (\ref{C1}) are:
cochlear length $3.5 \, cm$;
mass density $0.05 \, g/cm$;
$\gamma = 0.2$;
$\lam = 0.2 \, cm$; 
$N = 16.67 \, (dyne/cm^3)$; 
$r_0 = 0.001$ $g/(cm^2 ms)$.

The middle ear serves as a low pass filter 
from sound pressure at eardrum to the 
displacement of stapes, and as a high pass filter from the sound pressure 
at eardrum to the acceleration. 
We fit the data in \cite{Guin_Peake_66} with the 
following gain factor $a_{M}(\omega)$ for each input $e^{i\omega \, t}$:
\be
a_{M}(\omega) = 30 ({1\over 30} + 0.0605\omega^2 
((1- {\omega^{2}\over \omega_{m}^{2}})^{2}+ 
(2\xi_{m}\omega /\omega_{m})^{2})^{-1/2}),
\label{C23}
\ee
where $\omega_m = 4 \, kHz$, the middle ear characteristic frequency,  
and $\xi_{m}=0.7$ the middle ear damping ratio. 

\subsection{Dispersion Effect}
\begin{figure}[p]
\centerline{\includegraphics[width=350pt,height=340pt]{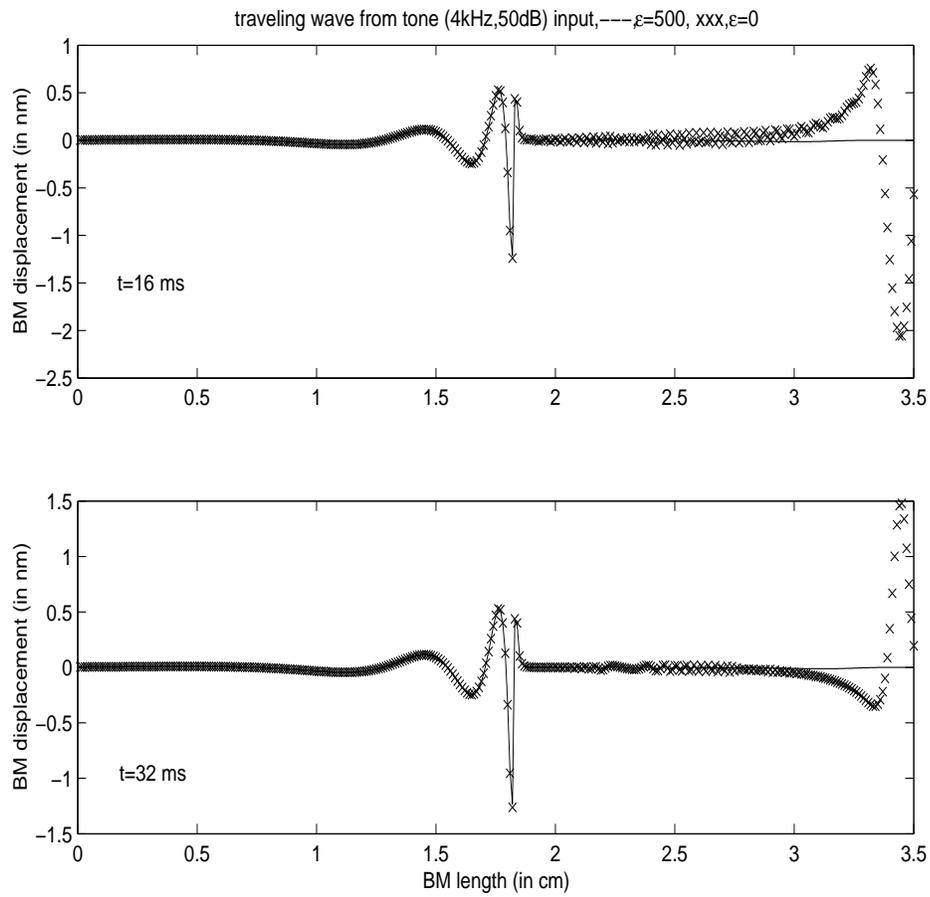}}
\caption{BM displacement at $t=16$ ms (top) and $t=32$ ms (bottom), 
from a (4kHz,50dB) input, 
with (line)  and without (cross) selective damping.}
\label{Fig1}
\end{figure}

We illustrate the dispersion property of time dependent solutions  
numerically in case of one tone input. 
In top plot of Fig. \ref{Fig1}, 
we show the profiles of
computed traveling wave at $t=16$ ms with and without the selective 
damping term, $\eps(x) \, u_t$. For the run, $h=0.01$, $k=0.01$, and 
$x_s =3$, $\beta=4$. We see that for a tone input of $4kHz$, $50dB$, without 
selective damping, the BM displacement at $t=16$ ms contains a long wave 
which manages to pass through the characteristic location in the interval $(1.5,2)$ 
and persists near the exit $x=L$. This is due to the dispersive long waves 
with slow decay as we discussed in the last section. 
In contrast, when the selective damping term 
is present, the long wave is damped as it moves 
close to the right end point $x=L$.
As the selective damping term is only effective near $x=L$, the solutions in 
the interior of the domain are essentially the same. In the bottom plot of 
Fig. \ref{Fig1}, we show the BM displacement at $t=32$ ms, the dispersive 
long wave persists. In other runs (not shown here), we also observed 
growth in amplitude 
of long waves near $x=L$ with time at a linear rate. The appearance of 
long waves seems to 
be independent of numerical methods, 
rather their existence is intrinsic to the model 
which is dispersive. 

\subsection{Numerical Parameters}
For all our runs reported below, 
$h=k=0.01$, $\eps =500$, $\beta =4$, $x_s=3$. The $x_s$ and $\beta$ 
values should be properly 
increased ($x_s$ closer to $x=3.5$) if there are low frequencies below 
$300 Hz$ in the input frequencies. 
Smaller $h$ and $k$ 
values have been used to test that numerical solutions 
do not vary much with 
further refinement. 
The time step $k$ should be made suitably smaller if input 
frequencies are as large as $16kHz$ for more resolution.

\subsection{Isodisplacement Curves}
For each input pressure wave of frequency $\omega $, there is 
a unique location $x_{\omega} \in [0,L]$ where the maximum 
of BM displacement (peak in absolute value) is located. The 
$x_{\omega} $ is   
called characteristic place. Conversely, for each location $x \in [0,L]$ 
there is a frequency, called characteristic frequency (CF) and 
denoted by $CF(x)$, such that the BM response amplitude attains 
the maximum at $x$. 
In Fig. \ref{Fig1}, we see that 
$x_\omega \in (1.5, 2)$ for $\omega = 4 $ kHz. The higher the amplitude 
of the input wave (denoted by $A_{in}$), 
the higher the BM response peak. If the level of 
BM response is set at a fixed level, such as 1 nm, and a BM location 
$x$ is given, we look for  
an input value $A_{in}$ so that $|u|(x) = 1$ nm. Here $|u|$ is the 
steady state BM response for input wave $A_{in}e^{i\omega t}$. 
The plot of $A_{in}$ as a function of $x$ gives the 
so called isodisplacement curve. The profile $|u|(x)$ is 
asymmetric about the peak, and decays rapidly to zero beyond 
$x_{\omega}$. So it takes higher $A_{in}$ to stimulate a point 
$x > x_{\omega}$ to 1 nm than to the left. It takes least 
$A_{in}$ to stimulate $|u|(x_\omega ) $ to 1 nm. So the 
isodisplacement curve has a minimum at $x_\omega$, rises 
sharply at $x > x_\omega $ and
gradually at $x < x_\omega $.     

\begin{figure}[p]
\centerline{\includegraphics[width=250pt,height=250pt]{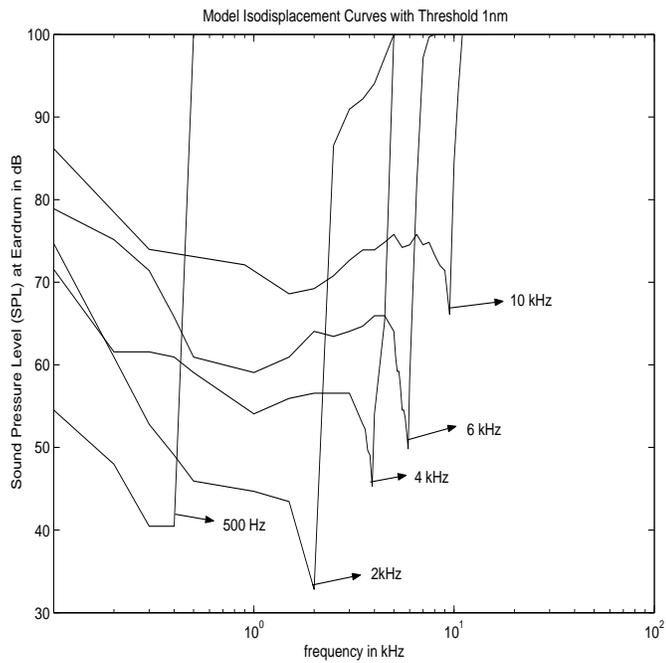}}
\caption{Isodisplacement curves (1 nm BM displacement) at five 
characteristic frequencies 500 Hz, 2 kHz, 4 kHz, 6 kHz, 10 kHz.}
\label{Fig3}
\end{figure}

In Fig. \ref{Fig3}, we show model isodisplacement curves for five 
characteristic frequencies 500 Hz, 2 kHz, 4 kHz, 6 kHz, 10 kHz. 
The threshold 
displacement is 1 nm. The curves showed the trend of sharper tips in high 
frequencies, wider tips for lower frequencies, and the asymmetry 
about the tips. The plot is comparable to the one in \cite{Neely_85}.
Isodisplacement curves are related to frequency selectivity of 
hearing.


\subsection{Two and Three Tone Interactions}
\begin{figure}[p]
\centerline{\includegraphics[width=350pt,height=340pt]{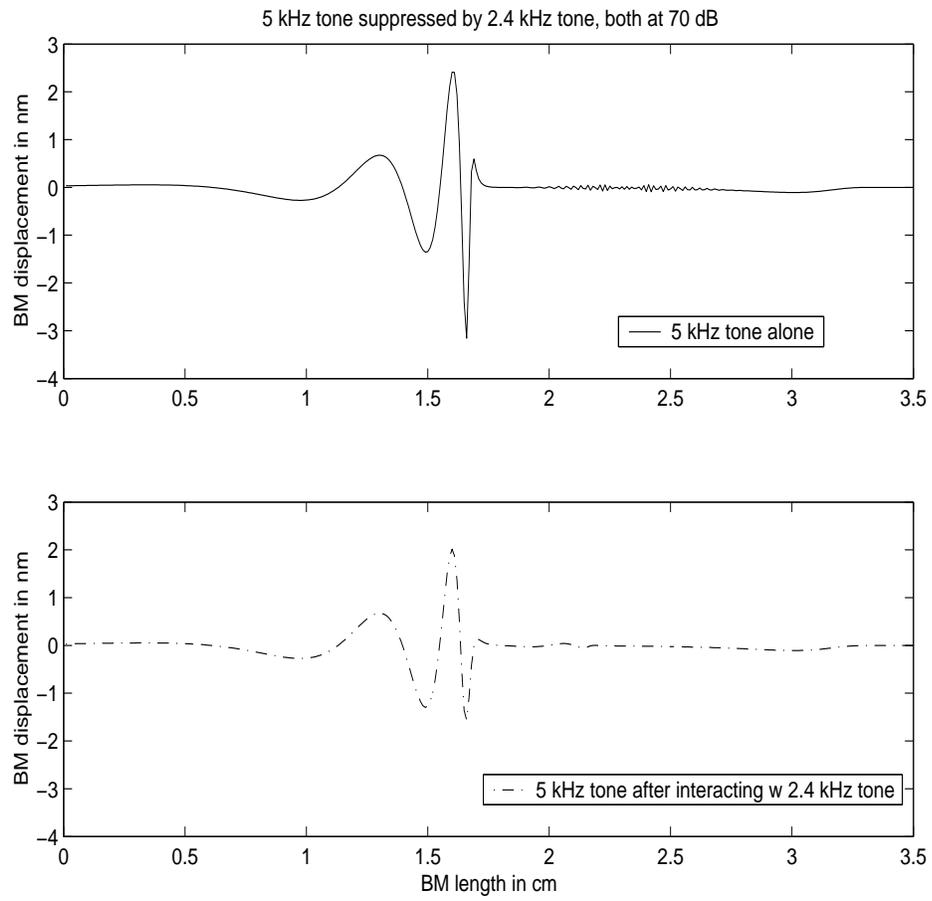}}
\caption{
Low side masking effect of 2.4 kHz tone on 5kHz tone both at 70 dB.
}
\label{Fig4}
\end{figure}

Two tone interaction is well-documented in the literature, 
here we illustrate 
that our model gives qualitatively the 
same results as shown in related figures in 
\cite{Geisler_98}. We shall use BM displacement to detect the 
tonal interaction. Fig. \ref{Fig4} shows that the amplitude of a 5 kHz 
70 dB tone drops (by nearly 50\%) after the second tone of 
2.4 kHz and 70dB is 
introduced, this reveals the so called low side suppression.

\begin{figure}[p]
\centerline{\includegraphics[width=350pt,height=340pt]{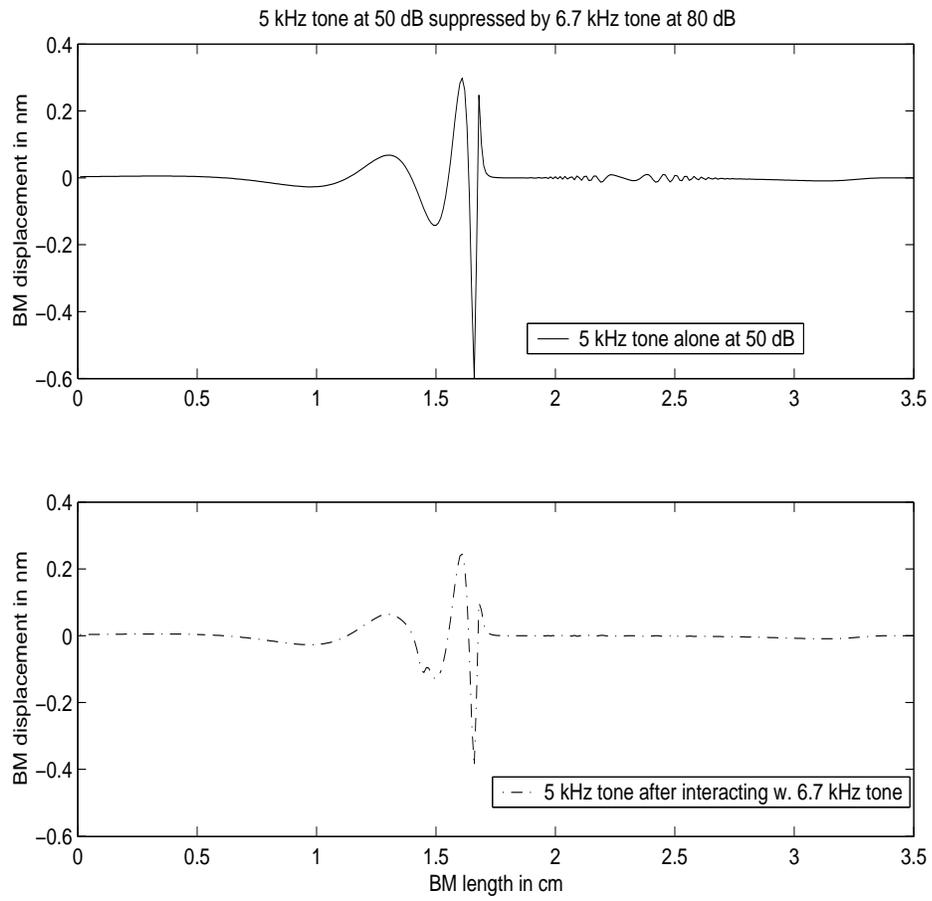}}
\caption{
High side masking effect of 6.7 kHz tone at 80 dB on 5kHz tone at 50 dB.}
\label{Fig6}
\end{figure}

\nit In Fig. \ref{Fig6}, we see that the 5 kHz tone at 50 dB (solid line)
is suppressed (by about 30 \%) after interacting 
with 6.7 kHz tone at 80 dB (plus-line), 
the so called high side suppression.
The high side suppressor tone at 6.7 kHz 
has to be much higher in amplitude than a
low side suppressor tone whose frequency is at least as far away from that 
of the suppressed tone. 
This shows the asymmetry of low and high 
side suppression as in \cite{Geisler_98} among others. 

\begin{figure}[p]
\centerline{\includegraphics[width=350pt,height=340pt]{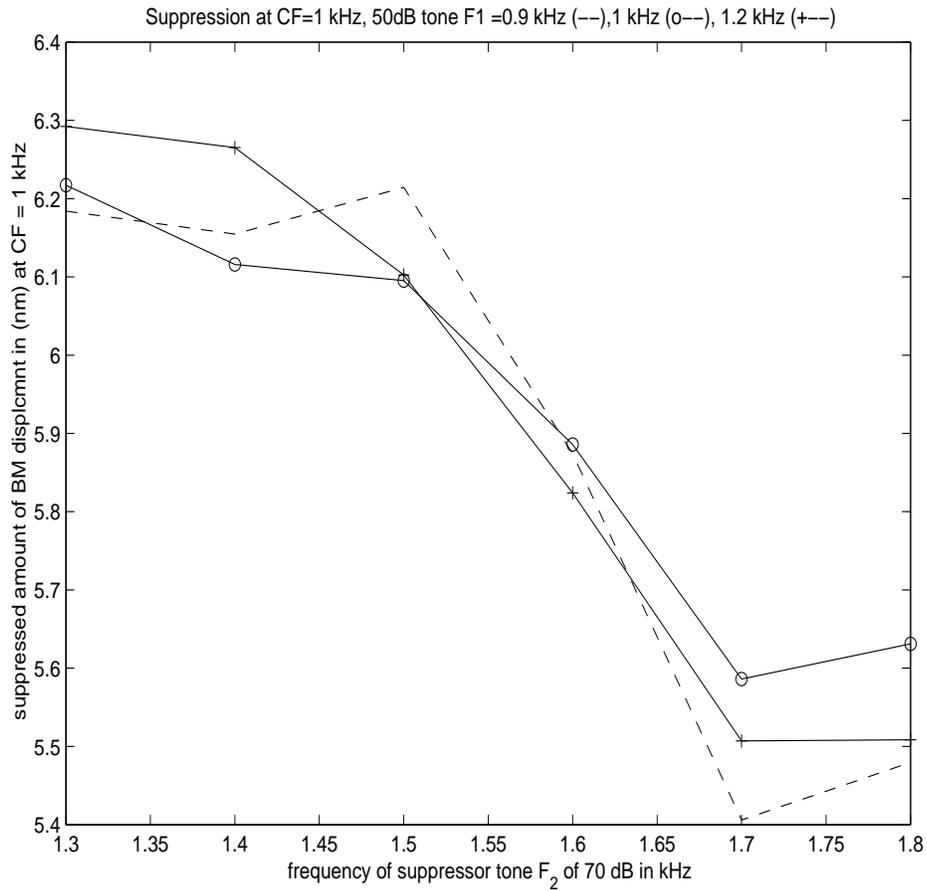}}
\caption{
The suppressed BM displacement (in nm) at characteristic frequency CF =1 kHz
in the presence of two tone $(F_1,F_2)$ input, for 50 dB
$F_1 = $ 0.9 kHz (dashed line), 1 kHz (circle-line), 1.2 kHz (plus-line), as
a function of $F_2$ which takes frequency values 
(1.3, 1.4, 1.5, 1.6, 1.7, 1.8) kHz at 70 dB.
}
\label{Fig7a}
\end{figure}

In \cite{Deng_85}, Fig. 7a showed
cat auditory neural rate response at a fixed characteristic
frequency (CF)
demonstrating suppression effect for two tone $(F_1,F_2)$
input, where $F_1$ at fixed intensity takes three frequency
values from below CF to above CF, for various values of 
the frequency $F_2$ at 70 dB. 
In Fig. \ref{Fig7a}, we showed a qualitatively similar plot
computed with our model, where CF=1 kHz,
50 dB $F_1$ takes three frequency values (0.9, 1, 1.2) kHz,
70 dB $F_2$ increases its frequencies as 1.3 kHz, 1.4 kHz, 1.5 kHz,
1.6 kHz, 1.7 kHz, 1.8 kHz.

\begin{figure}[p]
\centerline{\includegraphics[width=350pt,height=340pt]{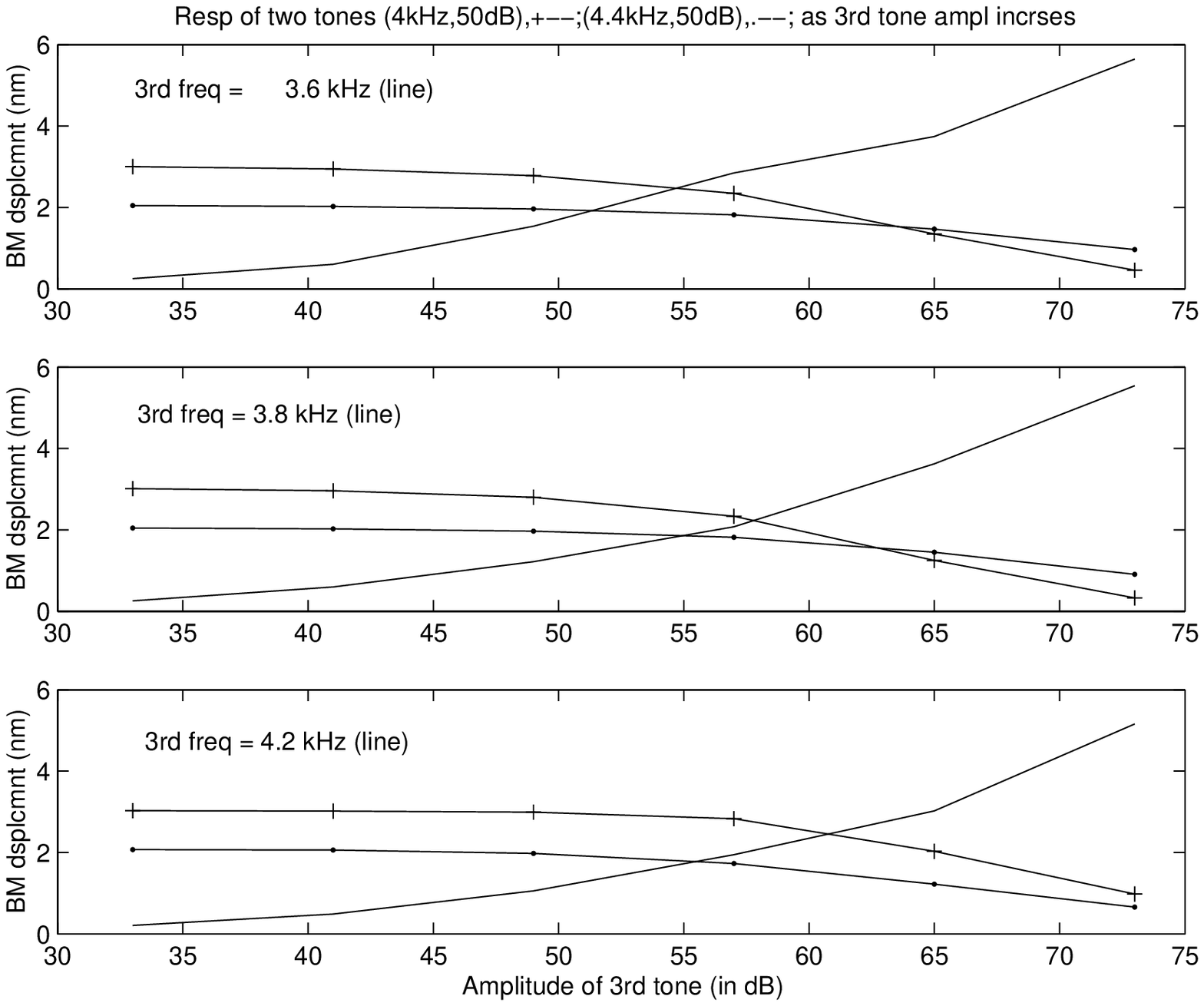}}
\caption{BM response of two 
tones 
(4 kHz, +--; 4.4 kHz, dot--) as the third tone (---) 
increases its amplitude from 33 dB to 73 dB at three values of
frequency, 3.6 kHz, 3.8 kHz, 4.2 kHz.
}
\label{Fig10}
\end{figure}
\begin{figure}[p]
\centerline{\includegraphics[width=350pt,height=340pt]{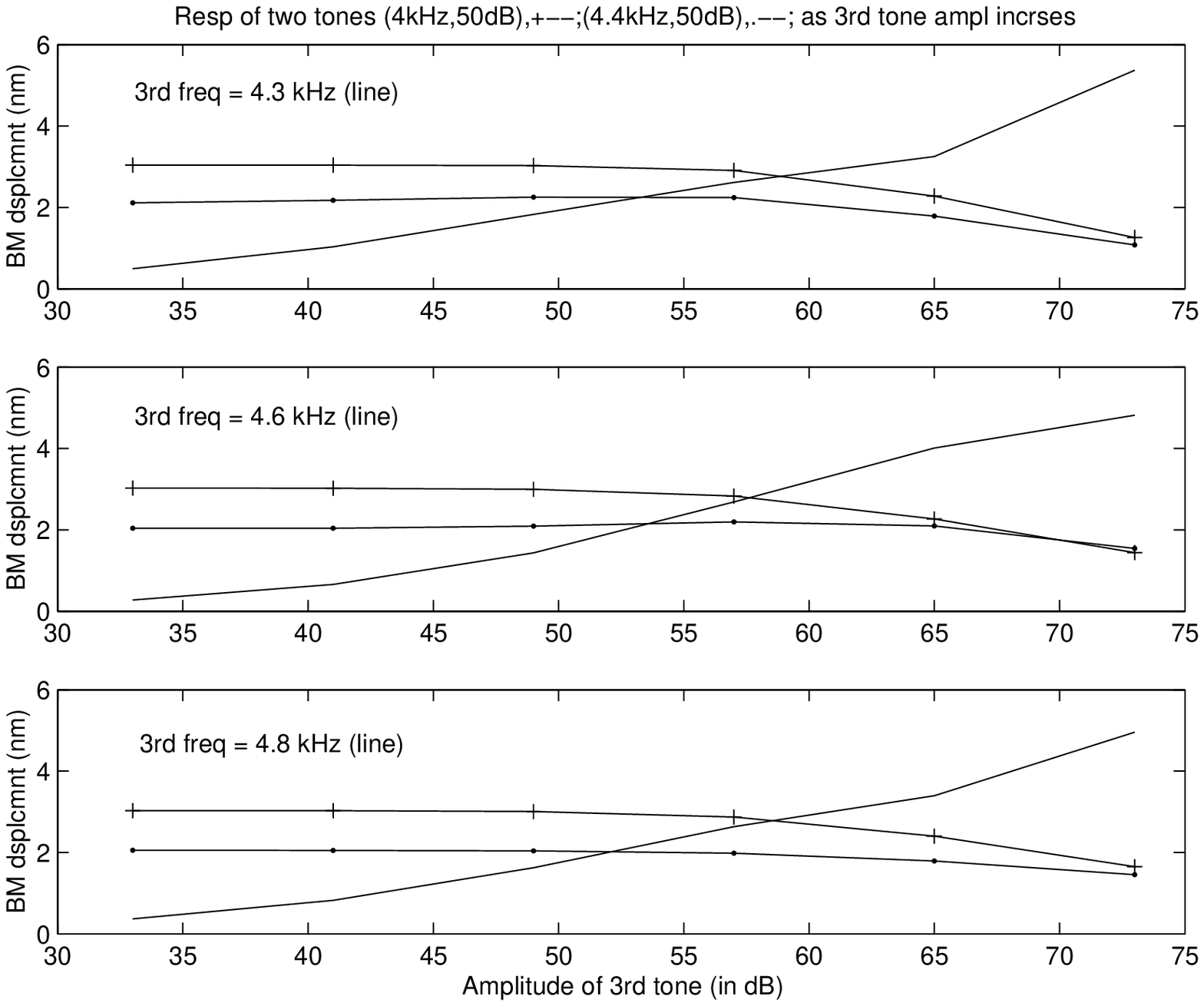}}
\caption{BM response of two
tones
(4 kHz, +--; 4.4 kHz, dot--) as the third tone (---)
increases its amplitude from 33 dB to 73 dB at three values of
frequency, 4.3 kHz, 4.6 kHz, 4.8 kHz.
}
\label{Fig11}
\end{figure}

The three tone interaction is much less documented in the literature, 
here we shall compare our model results qualitatively with the 
experimental data 
on auditory neural response to multi-tones 
in \cite{Deng_Geisler_87}. Two tones      
are fixed at 4 kHz, 4.4 kHz, both at 50 dB. 
The third tone varies in frequency and 
in amplitude. In Fig. \ref{Fig10}, we plot the BM response of two tones 
(4 kHz, plus line; 4.4 kHz, dotted line) as the third tone (line) 
increases its amplitude from 33 dB to 73 dB for three values of 
frequencies, 3.6 kHz, 3.8 kHz, 4.2 kHz.
We see that the two tone response decreases for increasing amplitude of third tone;
and that the suppression effect on the 4kHz tone is larger at 3.8 kHz (low side) 
than at 4.2 kHz (high side) for 
the same amplitude of third tone. The effect of third tone at 3.6 kHz is even 
larger than that at 4.2 kHz. In Fig. \ref{Fig11}, we plot similarly the 
response of two tones as the third tone takes on frequencies 4.3 kHz, 4.6 kHz, 4.8 kHz.
The masking effect of third tone remains, except that the second tone at 4.4 kHz 
shows more difference as we observe the crossing of two tone curves  
when the frequency of third tone goes through that of the second tone (4.4 kHz). 
The third tone is more efficient in masking the 
second tone from the low side (4.3 kHz) than from the high side (4.6 kHz). 
The suppressing effect on the second tone is even weaker at 4.8 kHz. 
These features agree with those of Fig. 3 of Deng and Geisler \cite{Deng_Geisler_87}, e.g. 
compare frame 4 and frame 6 there.
Strictly speaking
the neural data
in \cite{Deng_Geisler_87} are synchrony masking responses,
and BM responses we computed correspond to rate masking responses.
However, their data are relevant for qualitative comparison,
as there is close (positive) correlation between the rate and
phase behavior, see \cite{Deng_85}.

\section{Conclusions}
We discussed the dispersive property of the cochlear models for time 
domain computation, and slow decay of long waves. A selective 
longitudinal damping term has been introduced in the nonlinear nonlocal 
model studied previously \cite{Jau_Geisler_83,Deng_92}.
The new model is computed using a semi-implicit second order finite 
difference method in the time domain. We presented numerical results
on isodisplacement curves, two tone suppression, and three tone 
interaction in qualitative agreement 
with earlier findings of Geisler \cite{Geisler_98}
and experimental auditory neural data Deng and Geisler 
\cite{Deng_Geisler_87}. These encouraging results 
prompt one to perform further study of masking 
mechanism in more complex tones and speech input based on our cochlear 
model. 

\section{Acknowledgments}
J.X would like to thank Prof. C. D. Geisler for helpful email 
communications on two tone suppression and  
cochlear modeling. He wishes to thank Prof. J. Keller for his 
critical reading of the first manuscript and his constructive comments, 
and Prof. S. Osher for his encouragement. 
The authors wish to thank the Institute for
Mathematics and its Applications (IMA) at the University of Minnesota
for hosting the workshop on
speech processing in Sept of 2000 and for providing
an interactive scientific environment, where the initial collaboration
began. The authors also thank Mr. M. D. LaMar for his interest and 
careful reading of the paper. 
 
\vspace{.2 in}

\bibliographystyle{plain}

\begin{thebibliography}{99}
%
\bibitem{Allen_80}
        Allen, J. B.,
        {\em Cochlear Modeling--1980},
        in Lecture Notes in Biomathematics,
        Springer-Verlag,
        1980,
        eds. M. Holmes and L. Rubenfeld,
        volume 43,
        pp 1--8.   

\bibitem{Boer_96}
        de Boer, E.,
        {\em Mechanics of the Cochlear: Modeling Efforts},
        in Springer Handbook of Auditory Research,
        Springer-Verlag,
        1996,
        eds. P. Dollas and A. Popper and R. Fay,
        volume  8,
        pp 258--317.


\bibitem{Deng_92}L. Deng,
        {\em Processing of acoustic signals in a 
cochlear model incorporating 
laterally coupled suppresive elements},
        Neural Networks,
        5(1),
        pp 19--34,
        1992.

\bibitem{Deng_Geisler_87}L. Deng and C. D. Geisler,
        {\em Responses of auditory-nerve fibers to multiple-tone complexes},
        J. Acoust. Soc. Amer.,
        82(6),
        1989--2000,
         1987.

\bibitem{Deng_85}
        L. Deng and C. D. Geisler,
        {\em Changes in the phase of excitor-tone responses in 
cat auditory-nerve fibers by suppressor tones and fatique},
        J. Acoust. Soc. Amer,
        78(5),
        pp 1633--1643,
        1985.
    
\bibitem{Deng_K_93}
        L. Deng and I. Kheirallah,
        {\em Numerical property and efficient solution of a 
transmission-line 
model for basilar membrane wave motions},
        Signal Processing,
        volume 33,
        pp 269--285,
        1993.
     
\bibitem{Diep_etal_87}R. Diependaal, H. Duifhuis, H.W. Hoogstraten and 
M.A. Viergever,
        {\em Numerical methods for solving one-dimensional cochlear 
models in the time domain},
        J. Acoust. Soc. Amer,
        82(5),
        pp 1655--1666,
        1987.

\bibitem{Geisler_98}
        Geisler, C. D.,
        {\em From Sound to Synapse},
        Oxford University Press,
        Oxford,
        1998.  

\bibitem{Golub_Ort79}
        G. Golub and J. Ortega,
        {\em Scientific Computing and Differential Equations},
        Academic Press,
        San Diego, CA,
        1992.

\bibitem{Guin_Peake_66}
        J. J. Guinan and W. T. Peake,
        {\em Middle-Ear Characteristics of Anesthesized Cats},
        J. Acoust. Soc. Amer,
        41(5),
        pp 1237--1261,
        1967.

\bibitem{Hall_77}J. L. Hall,
        {\em Two-tone suppression in a nonlinear model 
of the basilar membrane},
        J. Acoust. Soc. Amer,
        volume 61,
        pp 802--810,
        1977.

\bibitem{Jau_Geisler_83}
        Y. Jau and C. D. Geisler,
        {\em Results from a cochlear model utilizing longitudinal coupling},
        in Mechanics of Hearing,
        Martinus Nijhoff Pub., Delft Univ. Press,
        1983,
        E. de Boer and M. Viergever,
        pp 169--176.      

\bibitem{Kanis_Boer_94}
        Luc-J. Kanis and E. de Boer,
        {\em Two-tone suppression in a locally active
nonlinear model of the cochlear},
        J. Acoust. Soc. Amer,
        96(4),
        pp 2156--2165,
        1994.

\bibitem{Keller_85} J. B. Keller, J. C. Neu,
{\em Asymptotic analysis of a viscous cochlear model }, 
J. Acoust. Soc. America, 77(6), pp 2107-2110, 1985. 

\bibitem{Kim_86}
        D. O. Kim,
        {\em An overview of nonlinear and active models},
        in Lecture Notes in Biomathematics: Peripheral Auditory Mechanisms,
        Springer-Verlag,
        1986,
        eds J. Allen and J.L. Hall and A. Hubbard and S.T. Neely and A. Tubis,
        volume 64,
        pp 239--249.

\bibitem{Peskin_88}R. Leveque, Ch. Peskin, P. Lax,
{\em Solution of a Two-Dimensional Cochlear Model with Fluid Viscosity},
SIAM J. Applied Math, v. 48, no. 1, pp 191-213, 1988. 

\bibitem{Lib_82}M. C. Liberman,
{\em The cochlear frequency map for the cat: Labeling auditory nerve fibers 
of known characteristic frequency},
        J. Acoust. Soc. Amer,
        volume 72,
        pp 1441--1449,
        1982.

\bibitem{Light_81}Lighthill, J.,
        {\em Energy flow in the cochlea},
        J. Fluid Mechanics,
        volume 106,
        pp 149--213,
        1981.


\bibitem{Lim_00}K-M Lim,
        {\em Physical and Mathematical Cochlear Models},
        PhD Dissertation, Department of Mechanical Engineering,
        Stanford University,
        2000.

\bibitem{Lut_86}B. Lutkenhoner and D. Jager,
      {\em Stability of active cochlear models: 
need for a second tuned structure},
      in Lecture Notes in Biomathematics: Peripheral Auditory Mechanisms,
        Springer-Verlag,
        1986,
        eds. J. Allen, J.L. Hall, A. Hubbard, S.T. Neely and A. Tubis,
        volume 64,
        pp 205--212.

\bibitem{Lyon_82}R. Lyon,
 {\em A computational model of filtering, detection, and compression in the 
cochlear}, IEEE International Conference on Acoustics, 
Speech and Signal Processing,
        pp 1282--1285,
        1982.

\bibitem{Meddis_01} R. Meddis, L. O'Mard, E. Lopez-Poveda,
{\em A computational algorithm for computing nonlinear auditory frequency 
selectivity}, J. Acoust. Soc. America 109(6), 2001, pp 2852-2861.

\bibitem{Neely_85}S. Neely,
{\em Mathematical modeling of cochlear mechanics}, 
        J. Acoust. Soc. Amer,
        78(1),
        pp 345--352,
        1985.

\bibitem{Osher_90} S. Osher, L. Rudin,
{\em Feature-Oriented Image Enhancement Using Shock Filters}, 
SIAM J. Numer. Analysis, Vol. 27, No. 4, pp 919-940, 1990. 

\bibitem{Rhode_71}W. S. Rhode,
{\em Observations of the vibration of the basilar membrane in the
squirrel monkey using the Moessbauer techniques},
        J. Acoust. Soc. Amer,
        volume 49,
        pp 1218--1231,
        1971.


\bibitem{Sondhi_80}Sondhi, M.,
        {\em The Acoustical Inverse Problem for the Cochlear},
        Lecture Notes in Biomathematics,
        Springer-Verlag,
        1980,
        eds M. Holmes and L. Rubenfeld,
        volume 43,
        pp 95--104. 

\bibitem{Strik_89}
        J. Strikwerda,
 {\em Finite Difference Schemes and Partial Differential Equations},
        Wadsworth and Brooks,
        Pacific Grove, CA,
        1989.

\bibitem{Bekesy_60}
        von B\'ek\'esy, G.,
        {\em Experiments in Hearing},
        McGraw-Hill,
        New York,
        1960. 

\bibitem{Wada_93}
H. Wada, K. Ohyama, S. Noguchi and T. Takasaka,
{\em Generation mechanism of toneburst-evoked otoacoustic emissions: 
	theoretical study},
        in Biophysics of Hair-Cell Sensory Systems,
        World Scientific, Singapore,
        1993,
        eds H.Duifhuis, J.Horst, P.van Dijk, S.M. van Netten,
        pp 94--101.

\bibitem{Whith_79}
        G. B. Whitham,
        {\em Linear and Nonlinear Waves},
        Wiley and Sons,
        New York,
        1979.    

\end{thebibliography}

\newpage

\newpage


\newpage

\end{document}